\documentclass[10pt,a4paper]{article}

\usepackage[left=2cm,right=2cm,top=2cm,bottom=2cm]{geometry}
\usepackage[english]{babel}
\usepackage{amsmath}
\usepackage{amsfonts}
\usepackage{amssymb}
\usepackage{bbm}
\usepackage{latexsym}
\usepackage{lmodern}
\usepackage{mathrsfs}
\usepackage{slashed}
\usepackage{braket}
\usepackage{tensor}
\usepackage{simplewick}
\usepackage{array}
\usepackage{graphicx}
\usepackage{xcolor}
\usepackage[font=small,labelfont=it]{caption}
\usepackage[font=scriptsize]{caption}
\usepackage[hidelinks]{hyperref}
\hypersetup{
        colorlinks=true,
        linkcolor=blue,
        citecolor=magenta,
}
\usepackage[nottoc]{tocbibind} 

\def\C{\mathbb C}

\def\NN{\mathbb N}

\newcommand{\de}{\delta}

\renewcommand{\th}{\theta}
\newcommand{\ka}{\kappa}

\newcommand{\by}{\bar{y}}
\newcommand{\bt}{\bar{\theta}}
\newcommand{\bps}{\bar{\psi}}

\newcommand{\sfrac}[2]{{\textstyle\frac{#1}{#2}}}

\newcommand{\pa}{\partial}
\newcommand{\im}{{\mathrm{i}}}

\newcommand{\diff}{{\mathrm{d}}}

\newcommand{\unity}{\mathbbm{1}}
\newcommand{\dott}{{\boldsymbol{\cdot}}}

\newcommand{\beq}{\begin{equation}}
\newcommand{\eeq}{\end{equation}}
\newcommand{\eq}{\end{equation}}
\newcommand{\bea}{\begin{eqnarray}}
\newcommand{\eea}{\end{eqnarray}}
\newcommand{\with}{{\quad{\rm with}\quad}}
\newcommand{\for}{{\quad{\rm for}\quad}}
\renewcommand{\and}{{\quad{\rm and}\quad}}
\newcommand{\und}{{\qquad{\rm and}\qquad}}
\renewcommand{\=}{\ =\ }

\newcommand{\nn}{\nonumber}

\allowdisplaybreaks  
\DeclareMathOperator{\tr}{tr}

\newcommand{\ie}{{i.e.}}
\newcommand{\eg}{{e.g.}}

\newcommand{\Lagr}{ { \mathcal{L} } }
\newcommand{\Delt}{ { \mathcal{M} } }

\newcommand\Tstrut{\rule{0pt}{2.6ex}}         

\advance \textheight by \topskip
\makeatletter
\@addtoreset{equation}{section}

\makeatother
\usepackage{tikz}
\usetikzlibrary{positioning,arrows}
\usetikzlibrary{decorations.pathmorphing}
\usetikzlibrary{decorations.markings}
\tikzset{
particle/.style={thick,draw=black, postaction={decorate},
    decoration={markings,mark=at position .50 with {\arrow[line width=1pt]{>}}  }},
aparticle/.style={thick,draw=black, postaction={decorate},
    decoration={markings,mark=at position .5 with  {\arrow[line width=1pt]{<}}  }},
gluon/.style={decorate, draw=black,
    decoration={coil,aspect=0}},
scalar/.style={thick,dashed,draw=black}
 }

\begin{document}

\begin{titlepage}
\setcounter{page}{0}

\phantom{.}
\vskip 1.5cm

\begin{center}

{\LARGE \bf 
Quantum mechanical Nicolai maps \\[8pt]
on K\"ahler manifolds
}
\vspace{12mm}

{\Large Olaf Lechtenfeld \ and \ Fin M\"uller}
\\[8mm]
\noindent {\em
Institut f\"ur Theoretische Physik and Riemann Center for Geometry and Physics\\
Leibniz Universit\"at Hannover, Appelstrasse 2, 30167 Hannover, Germany}
\vspace{12mm}

\begin{abstract} 
\noindent
We prove that the classical part of the Nicolai map for a supersymmetric free
nonrelativistic particle moving on a K\"ahler manifold is linear in the K\"ahler metric 
and local for its velocity. Moreover, we give an exact expression for the one-loop
correction of the map in terms of the K\"ahler potential and point out a direct way 
to higher loop order. This provides another example for the rare cases of exact 
Nicolai maps, and the first all-order one-loop result. As a check we have verified the 
Nicolai-map conditions to zero and one-loop order (free-action and determinant-matching).
A graphical representation is given as well.
\end{abstract}

\end{center}

\end{titlepage}
%


\section{Supersymmetric quantum mechanics on a K\"ahler space}

\noindent
Supersymmetric quantum field theories in $d{=}1$ spacetime dimensions, \ie\ based on a time interval
parametrizing a worldline, have a long history (for a comprehensive review, see~\cite{tong}).
In particular, it is known since the early 1980s that supersymmetric particles living in a flat target space
and self-interacting via a superpotential~$W$ allow for an exact Nicolai map~\cite{Nic3}, which maps the
quantum-mechanical path-integral measure to that of the free theory (quadratic~$W$). 
Such transformations are very rare in higher base dimensions, where the Nicolai map always exists as a nonlocal 
and nonpolynomial field transformation~\cite{Nic1,Nic2}, to be constructed perturbatively in a coupling 
constant~\cite{FL,DL1,L1,L2}. The topic has enjoyed a recent revival~\cite{ALMNPP,LR1,LR3,Lprague}.
On the other hand, the target space for a supersymmetric particle may be some Riemannian manifold, which 
adds metric data to the theory and produces a four-fermi interaction proportional to the Riemann tensor. 
Also in this more general setting Nicolai maps~\footnote{
We speak of the map in plural because often they occur in families.}
exist, but they acquire quantum corrections, \ie\ become
a power series in~$\hbar$ on top of the perturbative expansion in one or more couplings~\cite{CLR,L3,L4}. 
The coupling parameters may parametrize a superpotential and/or the metric.

In this paper, we address the construction of Nicolai maps for a supersymmetric particle on 
a K\"ahler manifold, as a special case admitting ${\cal N}{=}\,4$ instead of the standard ${\cal N}{=}\,2$
supersymmetries. For simplicity, we omit any superpotential~$W$ although it could be added.
Our motivation is the detection of a new case of {\it exact\/} Nicolai maps, hinted at in~\cite{L4}.

We employ $N$ chiral and antichiral superfield coordinates in $d{=}1$, ${\cal N}{=}\,4$ superspace
$(t,\th_1,\th_2,\bt^1,\bt^2)$,
\begin{equation}
\begin{aligned}
Z^n(y_-,\th) &\= z^n(y_-) + \bps^n(y_-)\th_1 + \th_2\psi^n(y_-) + \th_1\th_2 F^n(y_-) \ ,\\
Z_n(y_+,\bt) &\= z_n(y_+) - \psi_n(y_+)\,\bt^1 - \bt^2\bps_n(y_+) - \bt^1\bt^2 F_n(y_+) \ ,
\end{aligned}
\end{equation}
with $n=1,2,\ldots,N$ and (anti-)chiral worldline parameters
\begin{equation}
y_\pm \= t \pm \tfrac{\im}{2}\th_i\bt^i 
\qquad\Rightarrow\qquad
D_i y = D_i \th_j = 0 \and
D^i \by = D^i \bt^j = 0\ .
\end{equation}
The Grassmann coordinates~$(\psi^n,\psi_n,\bps^n,\bps_n)$ are the superpartners of the target coordinates~$(z^n,z_n)$,
and $(F^n,F_n)$ denote auxiliary degrees of freedom, required to close the supersymmetry algebra off-shell.
The index position is important and distinguishes holomorphic from antiholomorphic coordinates, \eg\ $z^n{=}z_n^*$.
The Lagrangian for a supersymmetric particle of unit mass on a complex $N$-dimensional K\"ahler manifold with coordinates 
$(z^n,z_n)$ and K\"ahler potential $\mathrm{K}$ is given by the full superspace integral
\begin{equation}
\Lagr \= \int\!\diff^2\th\,\diff^2\bt\ \mathrm{K}(Z^\dott,Z_\dott)
\end{equation}
which evaluates to~\cite{bertolini}
\begin{equation} \label{Lagoff}
\begin{aligned}
\Lagr &\= K_m^n \bigl( \dot{z}^m\dot{z}_n + \im\,\bps^m\dot\psi_n + \im\,\bps_n\dot\psi^m + F^m F_n \bigr)
\ +\ K_{mn}^{k\ell}\,\bps^m\psi^n\bps_k \psi_\ell \\
&\quad + K_{mn}^k \bigl( \bps^m\psi^n F_k + \im\,\dot{z}^m\bps_k\psi^n \bigr)
\ +\ K^{mn}_k \bigr( \bps_m\psi_n F^k + \im\,\dot{z}_m\bps^k\psi_n \bigr)\ ,
\end{aligned}
\end{equation}
where we employed the abbreviations
\begin{equation}
K = \mathrm{K}(z^\dott,z_\dott) \und 
K_{mn\ldots}^{k\,\ell\ldots}(z^\dott,z_\dott) = 
\tfrac{\pa}{\pa z^m}\tfrac{\pa}{\pa z^n}\cdots\tfrac{\pa}{\pa z_k}\tfrac{\pa}{\pa z_\ell}\cdots K(z^\dott,z_\dott) \ .
\end{equation}

The action of the four supersymmetries on the fundamental fields is given in Table~1.
\begin{table}[h!]
\begin{center}
\begin{tabular}{|l|cccc|}
\hline
& $\de^1$ & $\de^2$ & $\de_1$ & $\de_2$ \\
\hline 
$z^n$ \Tstrut & $\bps^n$ & $\psi^n$ & $0$ & $0$ \\
$z_n$ & $0$ & $0$ & $\psi_n$ & $-\bps_n$ \\
$\psi^n$ & $F^n$ & $0$ & $0$ & $-\im\dot{z}^n$ \\
$\psi_n$ & $\im\dot{z}_n$ & $0$ & $0$ & $-F_n$ \\
$\bps^n$ & $0$ & $-F^n$ & $\im\dot{z}^n$ & $0$ \\
$\bps_n$ & $0$ & $-\im\dot{z}_n$ & $F_n$ & $0$ \\
$F^n$ & $0$ & $0$ & $\im\dot{\psi}^n$ & $\im\dot{\bps}^n$ \\
$F_n$ & $-\im\dot{\bps}_n$ & $\im\dot{\psi}_n$ & $0$ & $0$ \\
\hline
\end{tabular}
\caption{Supersymmetry transformations}
\end{center}
\end{table}
Each of these may be employed to express the Lagrangian~\eqref{Lagoff} as a supervariation,
so we are free to choose a linear combination
\begin{equation}
\Lagr+\tfrac{\diff}{\diff t}(\ldots) \= \ka^1\de^1\Delt^1 + \ka^2\de^2\Delt^2 + \ka_1\de_1\Delt_1 + \ka_2\de_2\Delt_2
\qquad\with\quad \ka^1+\ka^2+\ka_1+\ka_2=1\ .
\end{equation}
We make use of this R-symmetry ambiguity to pick the simple ``chiral'' choice
\begin{equation} \label{kachoice}
\ka^1=1 \and \ka^2=\ka_1=\ka_2=0
\end{equation}
and drop the index on $\de$ and~$\Delt$,
\begin{equation}
\Lagr+\tfrac{\diff}{\diff t}(\ldots) \= \de\,\Delt \quad\with\quad
\Delt\ = \int\!\diff\th_2\,\diff^2\bt\ \mathrm{K}(Z^\dott,Z_\dott)
\= K_m^n\bigl( \psi^m F_n - \im\,\dot{z}^m \psi_n \bigr) - K_k^{mn}\,\psi_m\bps_n\psi^k\ .
\end{equation}

\section{The flow operator}

\noindent
Our goal is to interpolate between our K\"ahler manifold given by~$K$ and flat space~$\C^N$.
To this end, we introduce a dimensionless K\"ahler parameter~$g$ defining a family
\begin{equation}
K(g;z^\dott,z_\dott) \qquad\with\quad 
K(1;z^\dott,z_\dott)= K(z^\dott,z_\dott) \und K(0;z^\dott,z_\dott)=z^n z_n\ .
\end{equation}
Next, we derive a functional differential equation for an arbitrary correlation function in the effective theory 
of~$(z^n,z_n)$ obtained by integrating out the auxiliary and fermionic degrees of freedom~$(F,\psi,\bps)$.
Employing the supersymmetry Ward identity one finds~\cite{L1}
\begin{equation} \label{flowder}
\begin{aligned}
\tfrac{\diff}{\diff g}\,\bigl\langle\,Y[z]\,\bigr\rangle_g^{F,\psi,\bps,z} 
&\= \bigl\langle\,\pa_g Y[z]\ +\ \im\,\smallint\!\diff t\ \pa_g\Lagr(t)\ Y[z]\,\bigr\rangle_g^{F,\psi,\bps,z} \\
&\= \bigl\langle\,\pa_g Y[z]\ +\ \im\,\smallint\!\diff t\ \pa_g\,\de\Delt(t)\ Y[z]\,\bigr\rangle_g^{F,\psi,\bps,z} \\
&\= \bigl\langle\,\pa_g Y[z]\ +\ \im\,\smallint\!\diff t\ \de\,\pa_g\Delt(t)\ Y[z]\,\bigr\rangle_g^{F,\psi,\bps,z} \\
&\= \bigl\langle\,\pa_g Y[z]\ +\ \im\,\smallint\!\diff t\ \pa_g\Delt(t)\ \de\,Y[z]\,\bigr\rangle_g^{F,\psi,\bps,z} \\
&\= \bigl\langle\,\pa_g Y[z]\ +\ \im\,\smallint\!\diff t\ \pa_g\Delt(t)\ 
\smallint\!\diff t'\ \bps^i(t')\,\tfrac{\de}{\de z^i(t')}\,Y[z]\,\bigr\rangle_g^{F,\psi,\bps,z}
\end{aligned}
\end{equation}
with
\begin{equation}
\pa_g \Delt \= (\pa_g K_m^n)\bigl( \psi^m F_n - \im\,\dot{z}^m \psi_n \bigr) - (\pa_g K_k^{mn})\,\psi_m\bps_n\psi^k\ ,
\end{equation}
where we indicate by a superscript the degrees of freedom to be integrated over, 
and we denote the value~$g$ of the K\"ahler parameter in a subscript.
Due to unbroken supersymmetry we do not need to normalize the path integrals underlying the correlators.
As a consequence, only connected Feynman diagrams contribute.

At this stage we perform the auxiliary-field integration.
The metric factor~$K^n_m$ yields an inverse determinant for the functional measure, 
which contributes $\im\tr\ln(K^\dott_\dott)$ to the effective action.
Since $F$ appears quadratically in the action and linearly in the functionals of~\eqref{flowder},
it gets replaced by its algebraic classical solution,
\begin{equation}
F^m \ \to\  -(K^{-1})^m_n K^n_{k\ell}\,\bps^k \psi^\ell \und
F_n \ \to\  -(K^{-1})_n^m K_m^{k\ell}\,\bps_k \psi_\ell
\end{equation}
with $(K^{-1})_n^m K_m^\ell=\de_n^\ell$. Thus we obtain
\begin{equation}
\tfrac{\diff}{\diff g}\,\bigl\langle\,Y[z]\,\bigr\rangle_g^{\psi,\bps,z}
\= \bigl\langle\,\pa_g Y\ +\ \smallint\!\!\smallint\!\diff t\,\diff t'\ \bigl\{
(\pa_g K^n_m)\,\dot{z}^m \psi_n + \im\,\Gamma_m^{k\ell}\,\psi^m\bps_k\psi_\ell \bigr\}(t)\
\bps^i(t')\,\tfrac{\de}{\de z^i(t')}\,Y\,\bigr\rangle_g^{\psi,\bps,z}
\end{equation}
with the shorthand
\begin{equation}
\Gamma_m^{k\ell} \= (\pa_g K_m^{k\ell}) - (\pa_g K_m^p)(K^{-1})_p^q\,K_q^{k\ell}\ .
\end{equation}
The functional average $\langle\ldots\rangle_g^{\psi,\bps,z}$ is to be performed with respect to the 
on-shell supersymmetric Lagrangian
\begin{equation} \label{Lagon}
\Lagr \= K_m^n \bigl( \dot{z}^m\dot{z}_n + \im\,\bps^m\dot\psi_n + \im\,\bps_n\dot\psi^m \bigr)
\ +\ \im\,K_{mn}^k\,\dot{z}^m\bps_k\psi^n\ +\ \im\,K^{mn}_k\,\dot{z}_m\bps^k\psi_n
\ +\ R_{mn}^{k\ell}\,\bps^m\psi^n\bps_k \psi_\ell 
\end{equation}
involving the Riemann tensor $R_{mn}^{k\ell}$ as the four-fermion coupling.
To the local classical action $\smallint\!\diff t\,\Lagr$ we must add
the nonlocal quantum piece $\im\tr\ln(K^\dott_\dott)$ alluded to above.

Formally performing the $(\psi,\bps)$ path integrals
we arrive at a coupling-flow equation for the correlator~$\langle\,Y\,\rangle_g^z$ in the effective bosonic theory,
\begin{equation}
\tfrac{\diff}{\diff g}\,\bigl\langle\,Y[z]\,\bigr\rangle_g^z
\= \bigl\langle\,\pa_g Y[z]\ +\ R_g[z]\,Y[z]\,\bigr\rangle_g^z
\end{equation}
with the functional differential (``flow'') operator
\begin{equation} \label{Rop}
R_g[z] \= \smallint\!\!\smallint\!\diff t\,\diff t'\ \bigl\langle\,\bigl\{
(\pa_g K^n_m)\,\dot{z}^m \psi_n + \im\,\Gamma_m^{k\ell}\,\psi^m\bps_k\psi_\ell \bigr\}(t)\
\bps^i(t')\,\tfrac{\de}{\de z^i(t')}\,\bigr\rangle_g^{\psi,\bps}\ .
\end{equation}
The final functional average $\langle\ldots\rangle_g^z$ is based on an effective bosonic action
\begin{equation}
S_g[z] \= \sum_{r=0}^\infty \hbar^r\,S_g^{(r)}[z]
\end{equation}
with the classical and one-loop contributions
\begin{equation}
S_g^{(0)} \= \smallint\!\diff t\,K^n_m\dot{z}^m\dot{z}_n \ ,\qquad
S_g^{(1)} \= \im\tr\ln(K^n_m)
-\im\tr\ln\bigl(K^n_m\,\tfrac{\diff}{\diff t} + K_m^{nq} \dot{z}_q\bigr)
-\im\tr\ln\bigl(K^n_m\,\tfrac{\diff}{\diff t} + K^n_{mp} \dot{z}^p\bigr) \ ,
\end{equation}
and higher-loop terms being generated from the four-fermi interaction.

We point out that, due to our choice~\eqref{kachoice} of supersymmetry, the flow operator only acts on the
upper-case coordinates. In other words, lower-case coordinates are inert under a change in the K\"ahler parameter.
The fermionic averaging in the expression~\eqref{Rop} generates a full two- and (partially fused) four-point function 
for $(\psi,\bps)$ in the K\"ahler geometry ($z$~background), which contributes a further (hidden) $g$~dependence.
To set up a power-series expansion we must therefore perturbatively compute
\begin{equation} \label{2p4p}
S_n^{\ i}(t,t') \= \bigl\langle\,\psi_n(t)\ \bps^i(t')\,\bigr\rangle_g^{\psi,\bps} \und
G^{m\ i}_{\ k\ell}(t,t') \= \bigl\langle\,\psi^m\bps_k\psi_\ell(t)\ \bps^i(t')\,\bigr\rangle_g^{\psi,\bps}\ ,
\end{equation}
based on the quadratic and quartic parts (in fermions) of the on-shell Lagrangian,
\begin{equation} \label{L2L4}
\Lagr_2 \= \im\,\bps^m\,\bigl( K^n_m\,\tfrac{\diff}{\diff t} + K_m^{nq} \dot{z}_q \bigr)\,\psi_n
+ \im\,\bps_n\,\bigl( K^n_m\,\tfrac{\diff}{\diff t} + K^n_{mp} \dot{z}^p \bigr)\,\psi^m
\und \Lagr_4 \= R_{mn}^{k\ell}\,\bps^m\psi^n\bps_k \psi_\ell\ .
\end{equation}

\section{The classical Nicolai map}

\noindent
The flow operator~$R_g$ embodies the infinitesimal Nicolai map, which therefore is obtained by iterated 
application or exponentiation. More concretely, given a power-series expansion of the flow operator,
\begin{equation}
R_g[z] \= r_1[z] + g\,r_2[z] + g^2 r_3[z] + g^3 r_4[z] + \ldots\ ,
\end{equation}
a universal formula exists~\cite{LR1} for the corresponding Nicolai map,
\begin{equation} \label{Tseries}
T_g z^k \= \!\sum_{\bf n} g^n\,c_{\bf n}\,r_{n_s}[z]\ldots r_{n_2}[z]\,r_{n_1}[z]\ z^k
\qquad\!\textrm{with}\quad
{\bf n} = (n_1,n_2,\ldots,n_s)\ ,\quad n_i\in\NN \ ,\quad {\textstyle\sum}_i n_i = n\ ,
\end{equation}
where $1\le s \le n$, the $n{=}0$ term is the identity, and the coefficients
\begin{equation} \label{cn}
c_{\bf n} \= (-1)^s \bigl[n_1\cdot(n_1{+}n_2)\cdots(n_1{+}n_2{+}\ldots+n_s)\bigr]^{-1}
\end{equation}
are related to the Stirling numbers of the second kind.
Writing out the first few terms and suppressing the functional argument~$z$ of $r_k$,
the perturbative Nicolai map reads~\cite{LR1}
\begin{equation} \label{Texp}
\begin{aligned}
T_g z^k &\= z^k \ -\ g\,r_1 z^k \ -\ \sfrac12g^2\bigl(r_2-r_1^2\bigr)z^k\ -\
\sfrac16g^3\bigl(2r_3-r_1r_2-2r_2r_1+r_1^3\bigr)z^k \\
&\quad -\sfrac{1}{24}g^4\bigl(6r_4-2r_1r_3-3r_2r_2+r_1^2r_2-6r_3r_1
+2r_1r_2r_1+3r_2r_1^2-r_1^4\bigr)z^k \ +\ O(g^5) \ ,
\end{aligned}
\end{equation}
while $T_g z_k = z_k$. An analogous formula exists for the inverse map.

So far we have suppressed the dependence on~$\hbar$, but let us revive it to trace the loop counting.
It turns out that the flow operator~\eqref{Rop} carries an overall factor of~$\hbar^{-1}$ (from the action)
and the fermionic $2p$-point functions in~\eqref{2p4p} are of the order~$\hbar^p$ and expansions in~$g\hbar$.
Therefore, we may split
\begin{equation}
r_1 = r_1^{(0)} + \hbar\,r_1^{(1)}\ ,\quad
r_2 = r_2^{(0)} + \hbar\,r_2^{(1)} + \hbar^2 r_2^{(2)}\ ,\quad
r_3 = r_3^{(0)} + \hbar\,r_3^{(1)} + \hbar^2 r_3^{(2)} + \hbar^3 r_3^{(3)}\ ,\quad
\textrm{etc.}\ ,
\end{equation}
and also
\begin{equation}
T_g z^k \= \sum_{r=0}^\infty \hbar^r\,T_g^{(r)} z^k\ .
\end{equation}

For the classical part of the map ($r{=}0$), we simply omit the four-fermi term in the flow operator~\eqref{Rop}
(containing~$\Gamma$) and disregard the fermion-loop contributions to the two-point function in~\eqref{2p4p},
writing $S^{(0)i}_{\ \ n}=S^{\ i}_n|_{\hbar=0}$.
The latter is then determined by the differential equation
\begin{equation} \label{S0eq}
\bigl( K^n_m\,\tfrac{\diff}{\diff t} + K_m^{nq} \dot{z}_q \bigr)\,S^{(0)i}_{\ \ n}(t,t') \= \de_m^i\,\delta(t{-}t')\ ,
\end{equation}
which we rewrite as
\begin{equation} 
\bigl( \de_m^n\tfrac{\diff}{\diff t} - K_{mp}^q\dot{z}^p(K^{-1})_q^n \bigr)\,K_n^\ell\,S^{(0)i}_{\ \ \ell}(t,t') 
\= \de_m^i\,\delta(t{-}t')
\end{equation}
to generate a perturbative expansion
\begin{equation}
\begin{aligned}
K_n^\ell\,S^{(0)i}_{\ \ \ell}(t,t') &\= C(t{-}t')\,\de_n^i
\ +\ \smallint\!\diff t_1\ C(t{-}t_1)\,A_n^{\ i}(t_1)\,C(t_1{-}t') \\
&\quad +\ \smallint\!\!\smallint\!\diff t_1\,\diff t_2\ C(t{-}t_1)\,A_n^{\ m}(t_1)\,C(t_1{-}t_2)\,A_m^{\ i}(t_2)\,C(t_2{-}t')
\ +\ \ldots
\end{aligned}
\end{equation}
with
\begin{equation}
\tfrac{\diff}{\diff t}\,C(t{-}t') \= \de(t{-}t')  \quad\und\quad
A_n^{\ i}(t) \= K_{n p}^q\dot{z}^p(K^{-1})_q^i(t) \ .
\end{equation}
Let us insert this expansion into the classical flow operator. Partially integrating $\pa_t$ we have
\begin{align}
R_g^{(0)}[z] &\= -\smallint\!\!\smallint\!\diff t\,\diff t'\, z^m(t)\,\pa_t\,(\pa_g K_m^n)(t)\,
S^{(0)i}_{\ \ n}(t,t')\,\tfrac{\de}{\de z^i(t')} \\
&\= -\smallint\!\!\smallint\!\diff t\,\diff t'\, z^m(t)\,\pa_t\,(\pa_g K_m^\ell)(K^{-1})_\ell^n(t)
\bigl\{ C(t{-}t') \de_n^i + \smallint\!\diff t_1\, C(t{-}t_1) A_n^i(t_1) C(t_1{-}t') + \ldots \bigr\}
\tfrac{\de}{\de z^i(t')}\ , \nn
\end{align}
where $\pa_t$ acts to the right on all functions of~$t$.

Let choose smooth coordinates around~$z{=}0$, \eg\ geodesic ones. For dimensional reasons, the K\"ahler potential then 
is a power-series expansion in $gz^\dott$ and $gz_\dott$ around flat space, hence
\begin{align}
K_m^n &\= \de_m^n + g\,(K_m^n)_1 + \tfrac12g^2(K_m^n)_2 + \,O(g^3z^3) \\
&\= \de_m^n + g\,\bigl[z^p (K_{mp}^n)_1 + z_q (K_m^{nq})_1\bigr] + 
\tfrac12g^2 \bigl[ z^p z^{p'}(K_{mpp'}^n)_2 + 2 z^p z_q(K_{mp}^{nq})_2 + z_q z_{q'}(K_m^{nqq'})_2\bigr] + \,O(g^3z^3)\ , \nn
\end{align}
where the $(\ldots)_s$ in the first line are homogeneous functions of~$z$, and in the second line they are numerical constants. 
We note that $(K_{np}^q)_0=0$.
Let us inspect the first two orders of the classical map. To this end we evaluate
\begin{equation} \label{r1r2}
\begin{aligned}
r_1^{(0)} &\= -\smallint\!\!\smallint\!\diff t\,\diff t'\, z^m(t)\,\pa_t (K_m^i)_1(t)\,C(t{-}t')\,\tfrac{\de}{\de z^i(t')}
\= \smallint\!\!\smallint\!\diff t\,\diff t'\, \dot{z}^m(t)\,(K_m^i)_1(t)\,C(t{-}t')\,\tfrac{\de}{\de z^i(t')}\ ,\\[4pt]
r_2^{(0)} &\= -\smallint\!\!\smallint\!\diff t\,\diff t'\, z^m(t)\,\pa_t \bigl\{(K_m^i)_2-(K_m^n)_1(K_n^i)_1\bigr\}(t)\,
C(t{-}t')\,\tfrac{\de}{\de z^i(t')} \\
&\qquad\! -\smallint\!\!\smallint\!\!\smallint\!\diff t\,\diff t'\,\diff t''\, z^m(t)\,\pa_t (K_m^j)_1(t)\,C(t{-}t')\,
(K_{jp}^i)_1 \dot{z}^p(t')\,C(t'{-}t'')\,\tfrac{\de}{\de z^i(t'')}\ .
\end{aligned}
\end{equation}
Iterating the action of $r_1^{(0)}$ on~$z^i$ yields
\begin{align} \label{r1r1}
r_1^{(0)} r_1^{(0)} z^i(t) &\= \smallint\!\!\smallint\!\!\smallint\!\diff t'\,\diff t''\,\diff \bar{t}\ 
z^m(t'')\,\pa_{t''} (K_m^j)_1(t'')\,C(t''{-}t')\,\tfrac{\de}{\de z^j(t')}\,z^p(\bar{t})\,\pa_{\bar{t}} (K_p^i)_1(\bar{t})\,C(\bar{t}{-}t)\\
&\= \smallint\!\!\smallint\,\diff t'\,\diff t''\  z^m(t'')\,\pa_{t''} (K_m^j)_1(t'')\,C(t''{-}t')\,\bigl\{ 
\de_j^p\,\pa_{t'}\,(K_p^i)_1(t') - \dot{z}^p(t')\, (K_{pj}^i)_1(t') \bigr\}\,C(t'{-}t) \ , \nn
\end{align}
where, for the functional variation of $(K_p^i)_1$, we employed the second variant of~$r_1^{(0)}$ above
(or partially integrated $\pa_{\bar t}$).
With this, the first two orders of the classical Nicolai map become
\begin{equation} \label{t1t2}
\begin{aligned}
-g\,r_1^{(0)} z^i(t) &\= g\,\smallint\!\diff t'\ z^m(t')\,\pa_{t'} (K_m^i)_1(t)\,C(t'{-}t') \qquad\und \\
-\tfrac12g^2(r_2^{(0)}-r_1^{(0)}r_1^{(0)})\,z^i(t) &\= \tfrac12g^2 \smallint\!\diff t'\,z^m(t')\,\pa_{t'}\, (K_m^i)_2(t')\,C(t'{-}t)\ .
\end{aligned}
\end{equation}
We emphasize the cancellations at the second order. In particular, all contributions carrying two free propagators
cancel out, leaving a single nonlocality~$C$ also here.

This type of cancellation is actually systematic. A careful but tedious iterative analysis of the higher-order terms 
reveals that in the combination
\begin{equation}
\ell\; r_k^{(0)}\,z^i\ -\ r_{k-\ell}^{(0)}\,r_\ell^{(0)}\,z^i \qquad\for\quad k>\ell = 1,2,3,\ldots
\end{equation}
only ``length-one'' terms survive, \ie~those with a single $C$~nonlocality. 
Since according to~\eqref{cn} both terms in this combination occur in~$T_g z^i$ with equal and opposite coeffcients,
the entire classical Nicolai map is built with a single free propagator~$C$ and thus {\it local for\/}~$\dot{z}^i$.
In fact, the obvious systematics in~\eqref{t1t2} holds true: the order-$g^n$ term of $T_g^{(0)}z^i(t)$ boils down to
\begin{equation}
\tfrac1{n!}\,\smallint\!\diff t'\,z^m(t')\,\pa_{t'}\, (K_m^i)_n(t')\,C(t'{-}t) \=
-\tfrac1{n!}\,\smallint\!\diff t'\,\dot{z}^m(t')\, (K_m^i)_n(t')\,C(t'{-}t)\ ,
\end{equation}
which implies the very simple classical map
\begin{equation} \label{T0}
\begin{aligned}
T_g^{(0)} z^i(t) &\= \smallint\!\diff t'\ z^m(t')\,\pa_{t'}\,K_m^i(t')\,C(t'{-}t) 
\= -\smallint\!\diff t'\ \dot{z}^m(t')\,K_m^i(t')\,C(t'{-}t) \\
&\= z^i(t) \ -\ \smallint\!\diff t'\ \dot{z}^m(t')\,(K_m^i-\de_m^i)(t')\,C(t'{-}t)
\end{aligned}
\end{equation}
or, equivalently, the local map
\begin{equation}
T_g^{(0)} \dot{z}^i(t) \= \tfrac{\diff}{\diff t}\,T_g^{(0)} z^i(t) \= K^i_m(t)\,\dot{z}^m(t)\ .
\end{equation}
The entire $g$ and $(z^\dott,z_\dott)$ dependence is packaged into the K\"ahler metric. It cannot be simpler than this!
Indeed, the same result follows from dimensional reduction of the chiral Nicolai map in~\cite{L4}.

As a double check, let us confirm the classical Nicolai-map requirement, the ``free-action condition''~\cite{Nic3,CLR}.
Because the free bosonic action is bilinear in~$(z^\dott,z_\dott)$ and $z_\dott$ is not transformed, this condition is
a linear one,
\begin{equation}
\smallint\!\diff t\ \de_i^n\,\bigl(T_g^{(0)}\dot{z}_n\bigr)\,\bigl(T_g^{(0)}\dot{z}^i\bigr) \=
\smallint\!\diff t\ \de_i^n\,\dot{z}_n\,T_g^{(0)}\dot{z}^i \=
\smallint\!\diff t\ K_m^n\,\dot{z}_n\,\dot{z}^m \ ,
\end{equation}
and it is obviously fulfilled since the right-hand side agrees with the bosonic part of~$\Lagr$ in~\eqref{Lagon}.
For the quantum Nicolai-map conditions we need to investigate the loop corrections~$T^{(r)}_g z^k$ for~$r{\ge}1$.

\section{The quantum Nicolai map}

\noindent
Let us now focus on the quantum corrections to the Nicolai map and begin with the leading order in~$\hbar$.
Its contribution to the flow operator reads
\begin{equation} \label{R1}
R_g^{(1)}[z] \= -\smallint\!\!\smallint\!\diff t\,\diff t'\, z^m(t)\,\pa_t\,(\pa_g K_m^n)(t)\,
S^{(1)i}_{\ \ n}(t,t')\,\tfrac{\de}{\de z^i(t')} + \im\smallint\!\!\smallint\!\diff t\,\diff t'\,
\Gamma_m^{k\ell}(t)\,G^{(1)m\ i}_{\quad\ k\ell}(t,t')\,\tfrac{\de}{\de z^i(t')}\ .
\end{equation}
The recursion relation
\begin{equation}
K_n^\ell(t)\,S^{\ i}_{\ell}(t,t') \= K_n^\ell(t)\,S^{(0)i}_{\ \ \ell}(t,t') + \im\smallint\!\diff\bar{t}\ 
K_n^\ell(t)\,S_\ell^{\ j}(t,\bar t)\,R_{jk}^{pq} (K^{-1})_q^r(\bar t)\,S^k_{\ p}(\bar t,\bar t)\,
K_r^m(\bar t)\,S^{(0)i}_{\ \ m}(\bar t,t')
\end{equation}
for the full fermionic propagator implies that 
\begin{equation}
K_n^\ell(t)\,S^{(1)i}_{\ \ \ell}(t,t') \= \im\smallint\!\diff\bar{t}\ K_n^\ell(t)\,S^{(0)j}_{\ \ \ell}(t,\bar t)\,R_{jk}^{pq} 
(K^{-1})_q^r(\bar t)\,S^{(0)k}_{\ \ \ \ \,p}(\bar t,\bar t)\,K_r^m(\bar t)\,S^{(0)i}_{\ \ m}(\bar t,t')\ ,
\end{equation}
which involves a tadpole of the classical fermion propagator
$S^{(0)k}_{\ \ \ \ \,p}=\langle\psi^k\;\bps_p\rangle_g^{\psi,\bps}|_{\hbar^0}$
(note the reversed index position).
The same is true for the (partially fused) four-point function at one loop,
\begin{equation}
G^{(1)m\ i}_{\quad\ k\ell}(t,t') \= S^{(0)m}_{\ \ \ \ \,k}(t,t)\,S^{(0)i}_{\ \ \ell}(t,t')\ .
\end{equation}

We investigate the perturbative expansion to order~$g^2$, which requires the flow operator to order~$g$.
Given that $R_{jk}^{pq}=O(g^2)$, only the second term in~\eqref{R1} is relevant. Recalling that
\begin{equation}
\Gamma_m^{k\ell} \= (\pa_g K_m^{k\ell}) - (\pa_g K_m^p)(K^{-1})_p^q\,K_q^{k\ell}
\= (K_m^{k\ell})_1 + g\,(K_m^{k\ell})_2 - g\,(K_m^p)_1(K_p^{k\ell})_1 + O(g^2)
\end{equation}
and
\begin{equation}
\begin{aligned}
K^m_p(t)\,S^{(0)p}_{\ \ \ \ \,k}(t,t)\, & K_\ell^q(t)\,S^{(0)i}_{\ \ q}(t,t') \= C(0)\de_m^k\,C(t{-}t')\de_\ell^i \\[2pt]
& + \smallint\!\diff\bar{t}\ \bigl\{ C(0)\de_m^k\,C(t{-}\bar t)\,A_\ell^{\ i}(\bar t)\,C(\bar t{-}t') +
C(t{-}\bar t)\,A^m_{\ k}(\bar t)\,C(\bar t{-}t)\,C(t{-}t')\de_\ell^i\bigr\} + O(g^2)
\end{aligned}
\end{equation}
with
\begin{equation} \label{Adef}
A_\ell^{\ i} \= K_{\ell p}^q \dot{z}^p (K^{-1})_q^i  \quad\und\quad
A^m_{\ k} \= K^{mq}_p \dot{z}_q (K^{-1})^p_k
\end{equation}
we arrive at
\begin{equation}
\begin{aligned}
r_1^{(1)} &\= \im\smallint\!\!\smallint\!\diff t\,\diff t'\ (K_m^{mi})_1(t)\,C(0)\,C(t{-}t')\,\tfrac{\de}{\de z^i(t')}\, \\[4pt]
r_2^{(1)} &\= \im\smallint\!\!\smallint\!\diff t\,\diff t'\ \bigl\{ (K_m^{mi})_2 - 2(K_p^{mi})_1(K_m^p)_1 - (K_m^{mp})_1(K_p^i)_1
\bigr\}(t)\, C(0)\,C(t{-}t')\,\tfrac{\de}{\de z^i(t')} \\
&\quad +\im\smallint\!\!\smallint\!\!\smallint\!\diff t\,\diff\bar t\,\diff t'\ 
(K_m^{m\ell})_1(t)\,C(0)\,C(t{-}\bar t)\,(K^i_{\ell p})_1\dot{z}^p(\bar t)\,C(\bar t{-}t')\,\tfrac{\de}{\de z^i(t')} \\
&\quad +\im\smallint\!\!\smallint\!\!\smallint\!\diff t\,\diff\bar t\,\diff t'\
(K_m^{\ell i})_1(t)\,C(t{-}\bar t)\,(K_\ell^{mq})_1\dot{z}_q(\bar t)\,C(\bar t{-}t)\,C(t{-}t')\,\tfrac{\de}{\de z^i(t')}\ .
\end{aligned}
\end{equation}
For the $O(\hbar\,g^2)$ part of the map, we must iterate
\begin{equation}
\begin{aligned}
r_1^{(1)} r_1^{(0)} z^i(t) &\= -\im\smallint\!\!\smallint\!\!\smallint\!\diff t'\,\diff t''\,\diff \bar{t}\ 
(K_m^{mj})_1(t'')\,C(0)\,C(t''{-}t')\,\tfrac{\de}{\de z^j(t')}\,z^p(\bar t)\,\pa_{\bar t}(K_p^i)_1(\bar t)\,C(\bar t{-}t) \\[2pt]
&\= -\im\smallint\!\!\smallint\!\diff t'\,\diff t''\ (K_m^{mj})_1(t'')\,C(0)\,C(t''{-}t')\,\bigl\{
\pa_{t'}(K_j^i)_1(t') - \dot{z}^p (K_{pj}^i)_1(t')\bigr\}\,C(t'{-}t)\ ,\\[4pt]
r_1^{(0)} r_1^{(1)} z^i(t) &\= \im \smallint\!\!\smallint\!\diff t'\,\diff t''\,\diff\bar t\  
\dot{z}^m(K_m^j)_1(t'')\,C(t''{-}t')\,\tfrac{\de}{\de z^j(t')}\,(K_p^{pi})_1(\bar t)\,C(0)\,C(\bar t{-}t) \\[2pt]
&\= \im\smallint\!\!\smallint\!\diff t'\,\diff t''\ \dot{z}^m(K_m^j)_1(t'')\,C(t''{-}t')\,(K_{pj}^{pi})_1(t')\,C(0)\,C(t'{-}t) \= 0\ ,
\end{aligned}
\end{equation}
because $K_{mp}^{nq}=O(g^2)$ implies $(K_{pj}^{pi})_1=0$, equivalently $(K^{pi}_i)_1$ is a numerical constant. 
With this, the first two orders of the leading quantum map read
\begin{equation}
\begin{aligned}
-g\,r_1^{(1)} z^i(t) &\= -\im g\,\smallint\!\diff t'\ (K_m^{mi})_1(t')\,C(0)\,C(t'{-}t) \qquad\und \\[4pt]
-\tfrac12g^2(r_2^{(1)}{-}r_1^{(1)}r_1^{(0)}{-}r_1^{(0)}r_1^{(1)})\,z^i(t) &\= -\tfrac{\im}2g^2 \smallint\!\diff t'\ 
\bigl\{ (K_m^{mi})_2 - 2(K_p^{mi})_1(K_m^p)_1 \bigr\}(t')\, C(0)\,C(t'{-}t) \\
&\qquad{-}\tfrac{\im}2g^2 \smallint\!\!\smallint\!\diff t'\,\diff t''\
(K_m^{\ell i})_1(t')\,C(t'{-}t'')\,(K_\ell^{mq})_1\dot{z}_q(t'')\,C(t''{-}t')\,C(t'{-}t)
\end{aligned}
\end{equation}
and again reveal cancellations: all of $r_1^2z$ cancel against part of $r_2z$, 
but this time a maximally nonlocal term (carrying three free propagators) survives.

Nevertheless, such cancellations generalize to higher orders in $g$ and~$\hbar$, in the combinations
\begin{equation}
\ell\; r_k^{(r)}\,z^i\ -\ r_{k-\ell}^{(r)}\,r_\ell^{(0)}\,z^i \quad\und\quad
r_k^{(r)}\,z^i\ -\ r_{k-1}^{(r-1)}\,r_1^{(1)}\,z^i\ ,
\end{equation}
which collapses any full classical fermion propagator plus tadpole insertions emanating 
from the external time~$t$ to a free one. As a consequence, $T_g^{(1)}\dot{z}^i(t)$ 
is given by one-loop $n$-point functions with all but one arguments integrated over.
Pulling all together, we obtain
\begin{equation} \label{T1}
\begin{aligned}
T_g^{(1)} z^i(t) &\= -\im\smallint\!\diff t'\,
\bigl\{ g\,(K_m^{mi})_1 + \tfrac12g^2(K_m^{mi})_2 - g^2(K_p^{mi})_1(K_m^p)_1 \bigr\}(t')\,C(0)\,C(t'{-}t) \\[2pt]
&\qquad{-}\tfrac{\im}2g^2 \smallint\!\!\smallint\!\diff t'\,\diff \bar{t}\
(K_m^{\ell i})_1(t')\,C(t'{-}\bar{t})\,(K_\ell^{mq})_1\dot{z}_q(\bar{t})\,C(\bar{t}{-}t')\,C(t'{-}t) +\ O(g^3) \\[4pt]
&\=  -\im\smallint\!\!\smallint\!\diff t'\,\diff \bar{t}\ K^{im}_p (K^{-1})^p_\ell(t')\,C(t'{-}\bar{t})\,\bigl\{ 
\de^\ell_m \de(\bar{t}{-}t') + \tfrac12 A^\ell_{\ m}(\bar{t})\,C(\bar{t}{-}t') + O(g^3) \bigr\}\,C(t'{-}t) \\[4pt]
&\ =:\  -\im\smallint\!\!\smallint\!\diff t'\,\diff \bar{t}\ K^{im}_p (K^{-1})^p_\ell(t')\,C(t'{-}\bar{t})\,
H^\ell_{\ m}(\bar{t},t')\,C(t'{-}t)
\end{aligned}
\end{equation}
with $A^\ell_{\ m}$ given by~\eqref{Adef}. This form suggests an all-order expression
\begin{align}
H^\ell_{\ m}(\bar{t},t') &\= \de^\ell_m \de(\bar{t}{-}t') + \tfrac12 (A^\ell_{\ m}\,C)(\bar{t},t')
+ \tfrac13 (A^\ell_{\ n}\,C\,A^n_{\ m}\,C)(\bar{t},t') + \tfrac14 (A^\ell_{\ n}\,C\,A^n_{\ k}\,C\,A^k_{\ \ell}\,C)(\bar{t},t')
+ \ldots \nn \\[4pt]
&\= -\bigl[Z^{-1} \ln(1-Z)\bigr]^\ell_{\ m}(\bar{t},t') \qquad\with\quad
Z^\ell_{\ m}(\bar{t},t') \= A^\ell_{\ m}(\bar{t})\,C(\bar{t}{-}t')\ ,
\end{align}
where we suppressed internal time arguments and integrations.

We can verify this result via the higher Nicolai-map conditions~\cite{CLR}
\begin{equation} \label{loopcond}
\smallint\!\diff t\ \de_i^n\,\dot{z}_n\,T_g^{(r)}\dot{z}^i \=
S_g^{(r)}{-}S_0^{(r)} + \im\tr\ln\bigl(\tfrac{\de T_g^{(r-1)}z}{\de z}\bigr)
\qquad\for\quad  r=1,2,\ldots\ ,
\end{equation}
where $S_g^{(r)}$ is the $r$-loop contribution to the effective $(z^\dott,z_\dott)$~theory, 
and the final term arises from the Jacobian determinant of the map at $r{-}1$~loops.
For $r{=}1$, we have the residual of the auxiliary-field integration plus 
the fermion determinant based on~\eqref{S0eq} and it complex conjugate,
\begin{equation}
\begin{aligned}
S_g^{(1)}{-}S_0^{(1)} &\=
\im\tr\ln(K^n_m)
-\im\tr\ln\bigl(K^n_m\,\tfrac{\diff}{\diff t} + K_m^{nq} \dot{z}_q\bigr)
-\im\tr\ln\bigl(K^n_m\,\tfrac{\diff}{\diff t} + K^n_{mp} \dot{z}^p\bigr) 
+2\im\tr\ln\bigl(\de_m^n\,\tfrac{\diff}{\diff t}\bigr)\\[4pt]
&\= -\im\tr\ln(K_m^n) 
-\im\tr\ln\bigl(\de_m^n + C\,(K^{-1})^n_\ell K^{\ell q}_m \dot{z}_q\bigr)
-\im\tr\ln\bigl(\de^n_m + C\,(K^{-1})^n_\ell K^\ell_{mp} \dot{z}^p\bigr)\\
&\= -\im\tr\ln(K_m^n) + \im\,\sum_{s=1}^\infty \tfrac{(-1)^s}{s} \bigl\{
\tr\bigl(C\,(K^{-1})^n_\ell K^{\ell q}_m \dot{z}_q\bigr)^s +
\tr\bigl(C\,(K^{-1})^n_\ell K^\ell_{mp} \dot{z}^p\bigr)^s \bigr\} \\
&\= -\im\tr\ln(K_m^n)
+2\im\sum_{s\ \textrm{even}} \tfrac{1}{s}\,
\tr\bigl(C\,(K^{-1})^n_\ell K^{\ell q}_m \dot{z}_q\bigr)^s\ ,
\end{aligned}
\end{equation}
because
\begin{equation}
\tfrac{\diff}{\diff t}(\ln K^\dott_\dott)^n_m \=
(K^{-1})^n_\ell\,K^\ell_{mp} \dot{z}^p\ +\ (K^{-1})^n_\ell\,K^{\ell q}_m \dot{z}_q
\end{equation}
implies that
\begin{equation}
\tr\bigl( \ldots C\,(K^{-1})^n_\ell K^\ell_{mp} \dot{z}^p\,C \ldots\bigr) \=
-\tr\bigl(\ldots C\,(K^{-1})^n_\ell K^{\ell q}_m \dot{z}_q\,C \ldots\bigr) 
\end{equation}
via partial integrations,
and hence the odd terms in the sum cancel while the even ones add up.
The Jacobian is triangular because $\tfrac{\de T_g z_\dott}{\de z^\dott}=0$.
Since $\tfrac{\de T_g z_\dott}{\de z_\dott}=\unity$ we only need
\begin{equation}
\frac{\de T_g^{(0)} z^i(t')}{\de z^k(t)} \=
\pa_t\bigl\{ K_k^i(t)\,C(t{-}t')\bigr\} - \dot{z}^p K^i_{pk}(t)\,C(t{-}t') \=
K_k^i(t)\,\de(t{-}t') + K_k^{iq} \dot{z}_q(t)\,C(t{-}t')\ ,
\end{equation}
which leads to
\begin{equation}
\begin{aligned}
\im\tr\ln\bigl(\tfrac{\de T_g^{(0)}z}{\de z}\bigr) &\=
\im\tr\ln(K_k^i) + \im\tr\ln\bigl(\de^i_k + (K^{-1})^i_\ell K^{\ell q}_k\dot{z}_q\,C \bigr) \\
&\= \im\tr\ln(K_k^i) - \im\,\sum_{s=1}^\infty \tfrac{(-1)^s}{s}\,
\tr\bigl((K^{-1})^i_\ell K^{\ell q}_k\dot{z}_q\,C\bigr)^s
\end{aligned}
\end{equation}
The right-hand side of~\eqref{loopcond} thus combines to
\begin{equation}
\textrm{RHS} \= \im\,\sum_{s=1}^\infty \tfrac{1}{s}\,\tr\bigl((K^{-1})^i_\ell K^{\ell q}_k\dot{z}_q\,C\bigr)^s\ .
\end{equation}
We note that the alternating sum from the Jacobian determinant has been turned into a non-alternating one.
On the other hand, inserting the time derivative of the one-loop map~\eqref{T1} into the left-hand side 
and employing $\pa_t C(t'{-}t)=-\de(t'{-}t)$ yields
\begin{equation}
\begin{aligned}
\textrm{LHS} &\= \im\smallint\!\!\smallint\!\diff t\,\diff\bar{t}\,\dot{z}_i\,\bigl\{
K^{im}_p (K^{-1})^p_\ell(t)\,C(t{-}\bar{t})\,H^\ell_{\ m}(\bar{t},t) \bigr\} \\[4pt]
&\= \im\,\sum_{s=1}^\infty \tfrac{1}{s}\,\tr\bigl( K^{mi}_\ell\dot{z}_i (K^{-1})^\ell_n\,C \bigr)^s 
\= \im\,\sum_{s=1}^\infty \tfrac{1}{s}\,\tr\bigl( C\,(K^{-1})^m_\ell K^{\ell q}_n\dot{z}_q \bigr)^s\ ,
\end{aligned}
\end{equation}
where in the last step we reversed the order of the symmetric ``matrices'' $K^{\dott i}_{\dott}\dot{z}_i$, 
$(K^{-1})^\dott_\dott$ and $C$ under the trace.
Since clearly RHS=LHS, we have indeed verified the determinant-matching condition to all orders in~$g$
and thereby confirmed the one-loop result
\begin{equation} \label{T1full}
T_g^{(1)} z^i(t) \= -\im\smallint\!\!\smallint\!\diff t'\,\diff \bar{t}\ K^{im}_p (K^{-1})^p_\ell(t')\,
C(t'{-}\bar{t})\, \sum_{s=1}^\infty \tfrac{1}{s}\,( A^\ell_{\ m}\,C)^{s-1}(\bar{t},t')\,C(t'{-}t)\ .
\end{equation}

Interestingly, the conditions~\eqref{loopcond} involve the map only linearly and therefore suggest 
a recursive construction of $T_g^{(r)}$ given $T_g^{(r-1)}$, after stripping off a factor of~$\dot{z}_i$
under the integral resulting from the right-hand side. This in fact leads directly to~\eqref{T1full}
but may not be unique because of potential total time derivatives under the left-hand integral. 
In any case, the higher-loop parts $T_g^{(r)}$ of the Nicolai map for $r>1$ promise to be more
complicated in view of the role of the four-fermi coupling. It remains to be seen whether all tree
parts still collapse due to cancellations in the universal formula. If so, then a graphical representation
of the map will look as follows.
\newcommand\scale{.70}
\begin{equation} \label{graphs}
\begin{aligned} 
T_g z^i(t) &\=
\
\scalebox{\scale}{
\begin{tikzpicture}[baseline={([yshift=-1ex]current bounding box.center)},thick]
\node at (0,0)[circle,fill,inner sep = 2pt] {};
\draw (1,0)--(0,0);
\node at (1.04,0) [circle, fill=white, inner sep=-8pt] {\Large $\otimes$};
\end{tikzpicture} 
} \\
&\quad +\ 
\hbar\ \left(
	\scalebox{\scale}{
	\begin{tikzpicture}[baseline={([yshift=-1ex]current bounding box.center)},thick]
	\node at (0,0)[circle,fill,inner sep = 2pt] {};
	\draw (0.9,0)--(0,0); 
	\draw (0.9,0)--(0.9-  0.75*0.288 , 0.75*1/2); 
	\draw (0.9,0)--(0.9+ 0.75*0.288 ,0.75*1/2); 
	\draw[dashed] (0.9+0.9,0)--(0.9,0); 
	\draw [domain=-30:180+30] plot ( { 0.75* cos(\x)/3 + 0.9  } , {   0.75* sin(\x)/3+0.75* 2/3 } );
	\phantom{
	\draw [domain=-30:180+30] plot ( { 0.75* cos(\x)/3 + 0.9  } , {-   0.75* sin(\x)/3-0.75* 2/3 } );
	}
	\end{tikzpicture} 
	}
  \ +\ \frac{1}{2}\ 
	\scalebox{\scale}{
	\begin{tikzpicture}[baseline={([yshift=-1ex]current bounding box.center)},thick]
	\node at (0,0)[circle,fill,inner sep = 2pt] {};
	\draw (0.7,0)--(0,0); 
	\draw (0.7,0)--(0.7-  0.65*0.288 , 0.65*1/2); 
	\draw (0.7,0)--(0.7+ 0.65*0.288 ,0.65*1/2); 
	\draw[dashed] (0.7+0.9,0)--(0.7,0); 
		\draw [domain=-30:180+30] plot ( { 0.65* cos(\x)/3+0.7  } , {   0.65* sin(\x)/3+0.65* 2/3 } );
		\draw[gluon] (0.7,0.65 ) -- (0.7,1.2);
	\phantom{
		\draw[gluon] (0.7+0.4  ,-1.2 ) --(0.7,-0.65 );
	}
	\end{tikzpicture} 
	}
		\ + \ 
	\frac{1}{3}\ 
	\scalebox{\scale}{
	\begin{tikzpicture}[baseline={([yshift=-1ex]current bounding box.center)},thick]
	\node at (0,0)[circle,fill,inner sep = 2pt] {};
	\draw (0.7,0)--(0,0); 
	\draw (0.7,0) -- ++(60:0.75); 
	\draw (0.7,0) -- ++(120:0.75);
	\draw (0.325, 0.65)--(1.075, 0.65);
	\draw[dashed] (0.7+0.9,0)--(0.7,0); 
		\draw[gluon] (0.7,0) ++(60:0.75) -- ++(45:0.55);
		\draw[gluon] (0.7,0) ++(120:0.75)-- ++(135:0.55);
	\phantom{
		\draw[gluon] (0.7+0.4  ,-1.2 ) --(0.7,-0.65 );
	}
	\end{tikzpicture} 
	}
\ +\ \cdots \ \right)
\\
&\quad +\ 
\hbar^2 
\left(
	\scalebox{\scale}{
	\begin{tikzpicture}[baseline={([yshift=-1ex]current bounding box.center)},thick]
	\node at (0,0)[circle,fill,inner sep = 2pt] {};
	\draw (1,0)--(0,0); 
	\draw (1,0)--(1-  0.65*0.288 , 0.65*1/2); 
	\draw (1,0)--(1+ 0.65*0.288 ,0.65*1/2); 
	\draw[dashed] (1+0.9,0)--(1,0); 
		\draw [domain=-30:180+30] plot ( { 0.65* cos(\x)/3+1  } , {   0.65* sin(\x)/3+0.65* 2/3 } );
		\draw (1,0.65)--(1-  0.65*0.288 , 0.65*1/2 + 0.65 ); 
		\draw (1,0.65)--(1+ 0.65*0.288 , 0.65*1/2 + 0.65 );  
		\draw [domain=-30:180+30] plot ( { 0.65* cos(\x)/3+1  } , {  0.65+ 0.65* sin(\x)/3+0.65* 2/3 } );
	\phantom{
		\draw [domain=-30:180+30] plot ( { 0.65* cos(\x)/3+1  } , {  - 0.65- 0.65* sin(\x)/3-0.65* 2/3 } );
	}
	\end{tikzpicture} 
	}
	\ +\ \frac{1}{2}\
	\scalebox{\scale}{
	\begin{tikzpicture}[baseline={([yshift=-1ex]current bounding box.center)},thick]
	\node at (0,0)[circle,fill,inner sep = 2pt] {};
	\draw (1,0)--(0,0); 
	\draw (1,0)--(1-  0.65*0.288 , 0.65*1/2); 
	\draw (1,0)--(1+ 0.65*0.288 ,0.65*1/2); 
	\draw[dashed] (1+0.9,0)--(1,0); 
		\draw [domain=-30:180+30] plot ( { 0.65* cos(\x)/3+1  } , {   0.65* sin(\x)/3+0.65* 2/3 } );
		\draw (1,0.65)--(1-  0.65*0.288 , 0.65*1/2 + 0.65 ); 
		\draw (1,0.65)--(1+ 0.65*0.288 , 0.65*1/2 + 0.65 );  
		\draw [domain=-30:180+30] plot ( { 0.65* cos(\x)/3+1  } , {  0.65+ 0.65* sin(\x)/3+0.65* 2/3 } );
		\draw[gluon] (1- 0.65*0.288 ,0.65*1/2) -- ++(135:0.55);
	\phantom{
		\draw [domain=-30:180+30] plot ( { 0.65* cos(\x)/3+1  } , {  - 0.65- 0.65* sin(\x)/3-0.65* 2/3 } );
	}
	\end{tikzpicture} 
	}
	\ +\ \frac{1}{2}\
	\scalebox{\scale}{
	\begin{tikzpicture}[baseline={([yshift=-1ex]current bounding box.center)},thick]
	\node at (0,0)[circle,fill,inner sep = 2pt] {};
	\draw (1,0)--(0,0); 
	\draw (1,0)--(1-  0.65*0.288 , 0.65*1/2); 
	\draw (1,0)--(1+ 0.65*0.288 ,0.65*1/2); 
	\draw[dashed] (1+0.9,0)--(1,0); 
		\draw [domain=-30:180+30] plot ( { 0.65* cos(\x)/3+1  } , {   0.65* sin(\x)/3+0.65* 2/3 } );
		\draw (1,0.65)--(1-  0.65*0.288 , 0.65*1/2 + 0.65 ); 
		\draw (1,0.65)--(1+ 0.65*0.288 , 0.65*1/2 + 0.65 );  
		\draw [domain=-30:180+30] plot ( { 0.65* cos(\x)/3+1  } , {  0.65+ 0.65* sin(\x)/3+0.65* 2/3 } );
		\draw[gluon] (1 ,1.3) -- ++(90:0.55);
	\phantom{
		\draw[gluon] (1 ,-1.3) -- ++(270:0.55);
	}
	\end{tikzpicture} 
	}
	\ +\ \cdots \right. \\[-10pt]
	&\qquad \qquad +\ \left.\scalebox{\scale}{
	\begin{tikzpicture}[baseline={([yshift=-1ex]current bounding box.center)},thick]
	\node at (-1/3,0)[circle,fill,inner sep = 2pt] {};
	\draw (1,0)--(-1/3,0);  
	\draw[dashed] (1+0.9,0)--(1,0); 
		\draw [domain=0:360] plot ( {   (  2 +  cos(\x)  )  /3} , {    sin(\x)  /3  } );
	\end{tikzpicture} 
	}
\ + \ \frac{1}{2}\ \scalebox{\scale}{
	\begin{tikzpicture}[baseline={([yshift=-1ex]current bounding box.center)},thick]
	\node at (-1/3,0)[circle,fill,inner sep = 2pt] {};
	\draw (1,0)--(-1/3,0); 
	\draw[dashed] (1+0.9,0)--(1,0); 
	\draw [domain=0:360] plot ( {   (  2 +  cos(\x)  )  /3} , {    sin(\x)  /3  } );
	\draw[gluon] (2/3 ,1/3) -- ++(90:0.55);
	\phantom{
		\draw[gluon] (2/3 ,-1/3) -- ++(270:0.55);
	}
	\end{tikzpicture} 
	}
\ + \ \frac{1}{4}\ \scalebox{\scale}{
	\begin{tikzpicture}[baseline={([yshift=-1ex]current bounding box.center)},thick]
	\node at (-1/3,0)[circle,fill,inner sep = 2pt] {};
	\draw (1,0)--(-1/3,0);  
	\draw[dashed] (1+0.9,0)--(1,0); 
	\draw [domain=0:360] plot ( {   (  2 +  cos(\x)  )  /3} , {    sin(\x)  /3  } );
	\draw[gluon] (2/3 ,1/3) -- ++(90:0.55);
	\draw[gluon] (2/3 ,-1/3) -- ++(270:0.55);
	\end{tikzpicture} 
	}
\ + \cdots 
\vphantom{
\scalebox{\scale}{
	\begin{tikzpicture}[baseline={([yshift=-1ex]current bounding box.center)},thick]
	\node at (0,0)[circle,fill,inner sep = 2pt] {};
	\draw (1,0)--(0,0); 
	\draw (1,0)--(1-  0.65*0.288 , 0.65*1/2); 
	\draw (1,0)--(1+ 0.65*0.288 ,0.65*1/2); 
	\draw[dashed] (1+0.9,0)--(1,0); 
		\draw [domain=-30:180+30] plot ( { 0.65* cos(\x)/3+1  } , {   0.65* sin(\x)/3+0.65* 2/3 } );
		\draw (1,0.65)--(1-  0.65*0.288 , 0.65*1/2 + 0.65 ); 
		\draw (1,0.65)--(1+ 0.65*0.288 , 0.65*1/2 + 0.65 );  
		\draw [domain=-30:180+30] plot ( { 0.65* cos(\x)/3+1  } , {  0.65+ 0.65* sin(\x)/3+0.65* 2/3 } );
		\draw[gluon] (1 ,1.3) -- ++(90:0.55);
	\phantom{
		\draw[gluon] (1 ,-1.3) -- ++(270:0.55);
	}
	\end{tikzpicture} 
	}	
}\right)
\\[-8pt]
&\quad +\ O(\hbar^3) 
\end{aligned}
\end{equation}
with the fat dots marking the external argument~$t$, solid lines standing for free fermion propagators~$C$ and
\begin{equation}
	\scalebox{\scale}{
\begin{tikzpicture}[baseline={([yshift=-1ex]current bounding box.center)},thick]
\node at (0,0) [circle, fill=white, inner sep=-8pt] {\Large $\otimes$};
\end{tikzpicture} 
	} \= K^i_m \dot{z}^m \quad,\qquad 
	\scalebox{\scale}{
\begin{tikzpicture}[baseline={([yshift=-1ex]current bounding box.center)},thick]
\draw[dashed] (0,0)--(0.9,0); 
\end{tikzpicture} 
	} \= K^{im}_p\,(K^{-1})^p_\ell \quad,\qquad 
	\scalebox{\scale}{
\begin{tikzpicture}[baseline={([yshift=-1ex]current bounding box.center)},thick]
\draw[gluon] (0,0)--(0.9,0); 
\end{tikzpicture} 
	} \= A^\ell_m \= K^{\ell q}_p \dot{z}_q\,(K^{-1})^p_m \quad.  
\end{equation}

\section{Conclusions}

\noindent
We have applied the canonical Nicolai-map construction to the quantum mechanical setting on a 
K\"ahler manifold, without an external potential. Such a map connects the quantum dynamics of a 
nonrelativistic particle in the K\"ahler space with the one in flat space. Taking advantage of the extended 
${\cal N}{=}\,4$ supersymmetry and its R-symmetry, we were able to find in closed form a chiral map 
to zero- and one-loop order, as an analytic expression in the K\"ahler metric. 
The classical map features a single nonlocality and could have been guessed from the free-action condition. 
At order~$\hbar$, the map is simpler than expected and given by the fermionic one-loop $n$-point functions. 
It fulfils the determinant-matching condition identically for any K\"ahler potential.
Higher-loop contributions to the Nicolai map might be extracted recursively, because they appear only linearly
in the higher-loop conditions thanks to the chiral nature of the map.
We propose a graphical representation of a conjectured simple form of the Nicolai map to any loop order.

In the second author's master's thesis, the bosonic two-point function has been evaluated for 
the case of the $\C P^N$ model using the inverse Nicolai map~\cite{mueller}. 
Systematic resummation of parts of the ``Nicolai perturbation'' series produced an infrared cutoff
in terms of the K\"ahler geometry.




\end{document}